\documentclass[12pt]{article} \textwidth=6in \oddsidemargin=0in
\usepackage{amssymb}
\begin{document}
\def\ov{\over} \def\be{\begin{equation}} \def\ee{\end{equation}}
\def\bc{\begin{center}} \def\ec{\end{center}} \def\noi{\noindent}
\def\s{\sigma} \def\x{\xi} \def\ld{\ldots} \def\cd{\cdots}
\def\({\Big(} \def\){\Big)} \def\s{\sigma} \def\t{\tau} \def\G{\Gamma} \def\inv{^{-1}}
\def\sp{\vspace{1ex}}  \def\iy{\infty} \def\ep{\varepsilon} \def\C{\mathcal C} \def\g{\gamma}\def\Z{\mathbb Z} \def\la{\lambda} \def\PP{\mathcal P} \def\P{\mathbb P} \def\e{\eta} \def\ph{\phi} \def\m{\mu} \def\z{\zeta} \def\I0t{\int_{\G_{0,\t}}}
\def\a{\alpha} \def\b{\beta} \def\ps{\psi} \def\l{\ell} \def\phy{\phi_\iy}
\def\sq{\sqrt{\s}} \def\sqp{\sqrt{\s'}} \def\PPb{\overline{\PP}}
 
\hfill November 30, 2017 

\bc{\bf\large Blocks and Gaps in the Asymmetric\\ \sp Simple Exclusion Process: Asymptotics}\ec

\begin{center}{\large\bf Craig A.~Tracy}\\
{\it Department of Mathematics \\
University of California\\
Davis, CA 95616, USA}\end{center}

\begin{center}{\large \bf Harold Widom}\\
{\it Department of Mathematics\\
University of California\\
Santa Cruz, CA 95064, USA}\end{center}

\begin{abstract}{In earlier work the authors obtained formulas for the probability in the asymmetric simple exclusion process that at time $t$ a particle is at site $x$ and is the beginning of a block of $L$ consecutive particles. Here we consider asymptotics. Specifically, for the KPZ regime with step initial condition, we determine the conditional probability (asymptotically as $t\to\iy$) that a particle is the beginning of an $L$-block, given that it is at site $x$ at time $t$. Using duality between occupied and unoccupied sites we obtain the analogous result for a gap of $G$ unoccupied sites between the particle at $x$ and the next one.}
\end{abstract}

\bc{\bf I. Introduction}\ec

The \textit{asymmetric simple exclusion process} (ASEP) on the integer lattice $\Z$ has remained an important stochastic model in nonequilibrium
statistical physics and interacting particle systems since its introduction by Frank Spitzer \cite{Sp} nearly fifty years ago. Nearly ten years ago, for the
case of \textit{step initial condition} (particles initially occupying the positive integer sites $\Z^+$), a formula for the distribution
 of the $m$th particle from the left \cite{tw3} was the starting point for the one-point probability distribution of the height function for the Kardar-Parisi-Zhang (KPZ)
 equation with narrow wedge initial conditions \cite{ACQ, SS}.  For a review
 of these developments in stochastic growth processes, see \cite{Co, QS}.

In \cite{tw4} the authors derived formulas associated with blocks of particles in ASEP analogous to those derived earlier for a single particle \cite{tw1,tw2}. First there was a formula for the probability  that at time $t$ the $m$th particle from the left is the beginning of a block of particles of length $L$ (or ``$L$-block'') starting at~$x$. Specifically it was for the probability $\PP_{L,Y}(x,m,t)$ of the event
\[x_m(t)=x,\ x_{m+1}(t)=x+1,\ldots,x_{m+L-1}(t)= x+L-1,\]
given the initial configuration $Y=\{y_1,\ld,y_N\}$. Here $x_m(t)$ denotes the position of the $m$th particle from the left at time $t$. The probability was given in two forms as sums of multiple integrals. One of these permitted an extension to infinite systems unbounded on the right. In the case of step initial condition ($Y=\Z^+$), the sum was shown to equal an integral involving a Fredholm determinant. 

Here we consider the following question for ASEP with step initial condition: what is the conditional probability (asymptotically as $t\to\iy$) that the $m$th particle from the left is the beginning of an $L$-block, given that it is at site $x$ at time $t$? Of course everything depends on how $x$ and $m$ depend on $t$. Our result is for the KPZ regime considered in \cite[\textsection5]{tw3}.

Recall that in ASEP particles jump one step to the right with probability $p$ (if the site is unoccupied) or one step to the left with probability $q=1-p$ (if the site is unoccupied). We assume $q>p>0$, so there is a drift to the left. The notation of \cite{tw3} was
\[m=\s t,\ \ c_1=-1+2\sqrt\s,\ \ c_2=\s^{-1/6}(1-\sqrt\s)^{2/3}.\]
What was shown there was that, with $\g=q-p$, 
\[\lim_{t\to\iy}\P(x_m(t/\g)\le c_1\,t+c_2\,s\,t^{1/3})=F_2(s),\]
uniformly for $\s$ in a compact subset of $(0,1)$, where $F_2$ is the distribution function of random matrix theory. Here we show the following, under the same assumptions. (We call this the KPZ regime.)

\noindent{\bf Theorem 1}. When $m=\s t$ and $x=c_1\,t+c_2\,s\,t^{1/3}$ we have, as $t\to\iy$,
\[\PP_{L,\,\Z^+}(x,m,t/\g)=c_2\inv\,\s^{(L-1)/2}\,{F_2}'(s)\,t^{-1/3}+o(t^{-1/3}).\]

\noindent{\bf Corollary 1}. The conditional probability that the $m$th particle from the left is the beginning of an $L$-block, given that it is at $x$ at time $t/\g$,  has the limit $\s^{(L-1)/2}$.\footnote{The conditional probability that there is a block of precisely $L$ particles, and no more, has the limit $\s^{(L-1)/2}-\s^{L/2}=\s^{(L-1)/2}\,(1-\sqrt\s)$.}

The corollary follows since the conditional probability is equal~to
\[{\PP_{L,\,\Z^+}(x,m,t/\g)\ov\PP_{1,\,\Z^+}(x,m,t/\g)}=\s^{(L-1)/2}+o(1).\]

We also consider the probability that there is a gap of (at least) $G$ unoccupied sites to the right of $x$. Specifically, we define $\PPb_{G,Y}(x,m,t)$ to be the probability that $x_m(t)=x$ and $x_{m+1}(t)>x+G$, given the initial configuration~$Y$. For this we show the following.

\noindent{\bf Theorem 2}. When $m=\s t$ and $x=c_1\,t+c_2\,s\,t^{1/3}$ we have, as $t\to\iy$,
\[\PPb_{G,\,\Z^+}(x,m,t/\g)=c_2\inv\,(1-\sq)^G\,{F_2}'(s)\,t^{-1/3}+o(t^{-1/3}).\]

\noindent{\bf Corollary 2}. The conditional probability that the $m$th particle from the left is followed by a gap of $G$ unoccupied sites, given that it is at $x$ at time $t/\g$, has the limit $(1-\sq)^G$.\footnote{The conditional probability that there is a gap of precisely $G$ sites, and no more, has the limit~$(1-\sq)^G\,\sq$. No gap is the same as a block of at least two, so this is consistent with Corollary 1 with $L=2$.}

The starting point for the proof of Theorem 1 will be the formula for $\PP_{L,\,\Z^+}(x,m,t/\g)$ derived in \cite{tw4}. We set $\t=p/q<1$, define
\[U(\x,\x')={p+q\x\x'-\x\ov \x'-\x},\ \ \ K_x(\x,\x')={\x^x\,e^{(p/\x+q\x-1)t}\ov p+q\x\x'-\x},\]
and denote by $K_{L,x}(z)$ the integral operator acting on functions on $\C_R$ with kernel
\[K_{L,\,x}(\x,\x';\,z)=q^{1-L}\,K_{x+L-1}(\x,\x')\,\prod_{j=1}^L U(z_j,\x).\]
(The notation is slightly different than in \cite{tw4}.) Here $\C_R$ is the circle with center zero and large radius $R$, depending on $\t$. The result~was the formula
\[\PP_{L,\,\Z^+}(x,m,t)=(-1)^{L-1}\,p^{L(L+1)/2}\,\t^{-(m-1)(L-1)}\]
\[\times\int_{\G_{0,\t}}\cd\int_{\G_{0,\t}}{1\ov z_1^L\,(qz_1-p)\,z_2^{L-1}\,(qz_2-p)\cd z_L\,(qz_L-p)}\,\prod_{i<j}{1\ov U(z_j,z_i)}\]
\be\times\,\left[\int{\det(I-\t^{-L}\,\la\,K_{L,\,x}(z))\ov (\la;\t)_m}\;{d\la\ov\la^L}\right]\,dz_L\cd dz_1.\label{PLZ}\ee
The $\la$-integration is over a contour enclosing the singularities of the integrand at $\t^{-j}$ for $j=0,\ld,m-1$.

We explain the notation. First, $(\la;\t)_m=\prod_{j=0}^{m-1}(1-\la\,\t^j)$. As for the iterated integral, $\G_{0,\t}$ is  a contour consisting of tiny circles around the points $z=0$ and $z=\t$, with the circles for each $z_{i}$ lying well inside the circles for $z_{i-1}$. Alternatively, the integral is interpreted as follows: First take the sum of the residues at $z_L=0$ and $z_L=\t$. In the resulting integrand take the sum of the residues at $z_{L-1}=0$ and $z_{L-1}=\t$. And so on.\footnote{The order  matters because of the factors $1/U(z_j,z_i)$ in the integrand. Observe that $U(\t,z)=p$ and $U(z,\t)=q$ for $z\ne\t$.}

The proof of the Theorem 1 will have two parts. In the first, a pair of facts on stability of Fredholm determinants allows us to replace the operators $K_{L,\,x}(z)$ in the $\la$-integral in (\ref{PLZ}) by different operators --- ones for which we can do an asymptotic analysis. Since the derivation of this replacement goes along the lines of the argument in \cite{tw3} for a single particle the details will be deferred to an appendix. In the second part we use the previously established formula to reduce the problem to the evaluation of an explicit $L$-dimensional integral (which is not completely trivial). Theorem 2 will be deduced from Theorem 1 using the duality between occupied and unoccupied sites in ASEP, and some easy computations.

\bc{\bf II. Replacing the operators \boldmath$K_{L,x}(z)$ by operators \boldmath$J_{L,x,m}(w)$}\ec

The details of what follows will be given in Appendix A. We first make the change of variables
\[\x={1-\t\e\ov 1-\e},\ \ \x'={1-\t\e'\ov 1-\e'}\]
in the operator, which will then act on functions on small circle about $\e=1$. Then, in the operator and the $z_i$-integrals in (\ref{PLZ}), we make the substitutions
\[z_i={w_i-\t\ov w_i-1},\]
and we find that the integrations are over $\G_{0,\t}$ as before, with the $w_i$-contours well inside the $w_{i-1}$ contours.

What comes next depends the two propositions, proved in \cite[\textsection2]{tw3}, on stability of Fredholm determinants.

\noi{\bf Proposition 1}. Suppose $r\to\C_r$ is a deformation of closed curves and a kernel $H(\e,\,\e')$ is analytic in a neighborhood of $\C_r\times\C_r\subset\mathbb{C}^2$ for each $r$. Then the Fredholm determinant of $H$ acting on $\C_r$ is independent of $r$.

\noi{\bf Proposition 2}. Suppose $H_1(\e,\,\e')$ and $H_2(\e,\,\e')$ are two kernels acting on a simple closed contour $\C$, that $H_1(\e,\,\e')$ extends analytically to $\e$ inside $\C$ {\bf or} to $\e'$ inside $\C$, and that $H_2(\e,\,\e')$ extends analytically to $\e$ inside $\C$ {\bf and} to $\e'$ inside $\C$. Then the Fredholm determinants of $H_1(\e,\,\e')+H_2(\e,\,\e')$ and $H_1(\e,\,\e')$ are equal.

After using these two propositions (among other things) we arrive at an operator $J_{L,x,m}(w)$ acting on functions on a circle with center zero and radius $r\in(\t,1)$. It has kernel
\be J_{L,x,m}(\e,\e';w)=\int{\ph_{\iy,x}(\z)\ov\ph_{\iy,x}(\e')}\,{\z^{m-L}\ov(\e')^{m-L+1}}\,{f(\m,\z/\e')\ov\z-\e}\,\prod_{j=1}^L V(\z,\e';w_j)\;d\z,\label{Jkernel}\ee
where
\[\ph_{\iy,x}(\e)=(1-\e)^{-x-L+1}\,e^{{\e\ov1-\e}t},\ \ \ f(\m,z)=\sum_{k\in\Z}{\t^k\ov1-\t^k\m}\,z^k,\ \ \ V(\z,\e';w)={w \,\z-\t\ov w \,\e'-\t}.\]
The $\z$-integration is over a circle with center zero and radius in the interval $(1,r/\t)$. The new statement, which will be derived in Appendix A, is
\[\PP_{L,\,\Z^+}(x,m,t)=-\t^{-(L^2-5L+2)/2}\int_{\G_{0,\t}}\cd\int_{\G_{0,\t}}
\prod_{j=1}^L{(w_j-1)^{L-j}\ov w_j(w_j-\t)^{L-j+1}}\,\prod_{i<j}{w_j-w_i\ov w_j-\t w_i}\]
\be\times\,\int\left[(\t^L\m;\t)_\iy\,\det(I+\m J_{L,x,m}(w))\,{d\m\ov\m^L}\right]\,dw_L\cd dw_1.\label{PJw}\ee
Here $\m$ runs over a circle of radius larger than $\t^{-L+1}$, and the order of integration of the $w_j$ is as indicated.  

\bc{\bf III. Asymptotics}\ec

Now we asssume $m=\s t$ and $x=c_1\,t+s\,c_2\,t^{1/3}$. In \cite[\textsection5]{tw3} we did a saddle point analysis of the operator with kernel (\ref{Jkernel}), but without the product in the integrand. We made the variable changes 
\be\e\to \x+c_3\inv\,t^{-1/3}\,\e,\ \ \ \e'\to\x+c_3\inv t^{-1/3}\,\e',\ \ \ \z\to \x+c_3\inv\,t^{-1/3}\,\z,\label{changes}\ee
where $\x=-\sqrt{\s}/(1-\sqrt{\s})$ was  the saddle point and $c_3=\s^{-1/6}\,(1-\sqrt{\s})^{5/3}$. Using Proposition 1 above, we found\footnote{There is a minor change here. In the expression for $\ph_{\iy,x}$ we have an exponent $-x-L+1$ rather than the $-x$ in \cite{tw3}. Since changing $x$ by $O(1)$ amounts to changing $s$ by $O(t^{-1/3})$ this does not affect the asymptotics.} that $\m$ times the operator had the same Fredholm determinant as an operator $J^{(0)}+o(1)$, where $o(1)$ denotes a family of operators whose trace norms tend to zero as $t\to\iy$. And $\det(I+J^{(0)})=F_2(s)$. The kernel of $J^{(0)}$ is
\be J^{(0)}(\e,\e')=\int_{\G_\z}{e^{-\z^3/3+s\z+(\e')^3/3-s\e'}\ov(\z-\e)(\e'-\z)}\,d\z,\label{J0}\ee
which is independent of $\m$. Here $\G_\z$ is the contour consisting of the rays from $-c_3$ to $-c_3+\iy\, e^{\pm2\pi i/3}$ while the operator acts on functions on $\G_\e$, which consists of the rays from $0$ to $\iy\,e^{\pm\pi i/3}$. 

Now consider the effect of the product in the integrand in (\ref{Jkernel}). If we make the  replacements (\ref{changes}) in $V(\z,\e';w)$ a little computation shows that
\[V(\z,\e';w)\to 1+(\z-\e')\,{w\ov w\,\x-\t}\,c_3\inv t^{-1/3}\,[1+O(\min(1,t^{-1/3}|\e'|))],\]
so for the product,
\be\prod_{j=1}^L V(\z,\e';w_j)\to 1+(\z-\e')\,\sum_{j=1}^L{w_j\ov w_j\,\x-\t}\,c_3\inv t^{-1/3}+E(\z,\e';w),\label{Vprod}\ee
where $E(\z,\e';w)$ is a polynomial in $\z-\e'$ with coefficients that are functions of $\e'$ with bound $O(t^{-2/3}|\e'|)$. By exactly the same argument as in \cite{tw3}, the error term $E(\z,\e';w)$ causing no difficulty, we see that $\m\,J_{L,x,m}(w)$ has the same Fredholm determinant as an operator 
\[J^{(0)}+o(1)+J^{(1)}\,\sum_{j=1}^L{w_j\ov w_j\,\x-\t}\,c_3\inv t^{-1/3}+o(t^{-1/3}),\]
where $J^{(1)}$ has kernel
\be J^{(1)}(\e,\e')=-\int{e^{-\z^3/3+s\z+(\e')^3/3-s\e'}\ov \z-\e}\,d\z.\footnote{This is obtained from the kernel of $J^{(0)}$ by multiplying the integrand in (\ref{J0}) by the factor $\z-\e'$ from (\ref{Vprod}).}\label{J1}\ee 
Both bounds are in trace norm, and the $o(1)$ bound is independent of $w$. 
Thus the determinant in (\ref{PJw}) is equal to the determinant of
\[I+J^{(0)}+o(1)+J^{(1)}\,\sum_{j=1}^L{w_j\ov w_j\,\x-\t}\,c_3\inv t^{-1/3}+o(t^{-1/3}),\]
which in turn equals 
\[\det(I+J^{(0)}+o(1))\,\det\(I+(I+J^{(0)})\inv J^{(1)}\,\sum_{j=1}^L{w_j\ov w_j\,\x-\t}\,c_3\inv t^{-1/3}+o(t^{-1/3})\)\]
\[=(F_2(s)+o(1))\ \Big[1+{\rm tr}\,((I+J^{(0)})\inv J^{(1)})\ \sum_{j=1}^L{w_j\ov w_j\,\x-\t}\,c_3\inv t^{-1/3}\Big]+o(t^{-1/3}).\]
Now from (\ref{J0}) and (\ref{J1}) we see that $J^{(1)}=dJ^{(0)}/ds$ and therefore by a general fact
\[{\rm tr}\,((I+J^{(0)})\inv J^{(1)})={d\ov ds}\log\det(I+J^{(0)})={{F_2}'(s)\ov F_2(s)},\]
so the determinant in (\ref{PJw}) is equal to
\be F_2(s)+o(1)+{F_2}'(s)\ \sum_{j=1}^L{w_j\ov w_j\,\x-\t}\,c_3\inv t^{-1/3}+o(t^{-1/3}).\label{deteq}\ee
Again the $o(1)$ term is independent of the $w_j$. 

The limit is independent of $\m$, so to evaluate the $\m$-integral in (\ref{PJw}) we will only need to use
\be\int (\t^L\m;\t)_\iy\ {d\m\ov\m^L}=(-1)^{L-1}{\t^{(L-1)(3L-2)/2}\ov(1-\t)\cd(1-\t^{L-1})}.\label{mint}\ee

Lastly we have to integrate over the $w_j$ in (\ref{PJw}), so it remains to evaluate
\be\int_{\G_{0,\t}}\cd\int_{\G_{0,\t}}F(w_1,\ld,w_L)\,dw_L\cd dw_1\label{F0int}\ee
and
\be\int_{\G_{0,\t}}\cd\int_{\G_{0,\t}}F(w_1,\ld,w_L)\,\sum_{j=1}^L{w_j\ov w_j\,\x-\t}\,dw_L\cd dw_1,\label{Fint}\ee
where
\[F(w_1,\ld,w_L)=\prod_{j=1}^L{(w_j-1)^{L-j}\ov w_j(w_j-\t)^{L-j+1}}\,\prod_{i<j}{w_j-w_i\ov w_j-\t w_i}.\]

\bc{\bf IV. End of the proof of Theorem 1}\ec

Write
\[F(w_1,\ld,w_L)={(w_1-1)^{L-1}\ov w_1(w_1-\t)^L}\,G_L(w_1,\ld,w_L),\]
where
\be G_L(w_1,\ld,w_L)=\prod_{j>1}{(w_j-1)^{L-j}\ov
w_j(w_j-\t)^{L-j+1}}\,\prod_{1\le i<j\le L}{w_i-w_j\ov\t w_i-w_j}.\label{G0}\ee
The integral (\ref{F0int}) is
\be\I0t\cd\I0t {(w_1-1)^{L-1}\ov w_1(w_1-\t)^L}\,G_L(w_1,\ld,w_K)\,dw_L\cd dw_1.\label{Gwint}\ee
It is shown in Appendix B that if $\ps(w_2,\ld,w_L)$ is analytic in the neighborhood of $\{0,\t\}^{L-1}$ then
\[\I0t\cd\I0t G_L(w_1,\ld,w_K)\,\ps(w_2,\ld,w_L)\,dw_L\cd dw_2\]
is analytic for $w_1$ outside $\{0,\t\}$ except for a pole of order at most $L-1$ at $w_1=1$, and is $O(1)$ for large $w_1$. If we use this with $\ps(w_2,\ld,w_L)=1$
we see by expanding the $w_1$-contour that the integral (\ref{Gwint}), which is the same as (\ref{F0int}), equals zero. (The pole of order $L-1$ is cancelled by the zero of order $L-1$ in the first factor.)
Similarly, by taking
\[\ps(w_2,\ld,w_L)=\sum_{j=2}^L{w_j\ov w_j\,\x-\t},\]
we see that the integral (\ref{Fint}), but with the sum starting at $j=2$, is also zero.

Thus the remaining integral to be evaluated is
\be\I0t\cd\I0t {(w_1-1)^{L-1}\ov w_1(w_1-\t)^L}\,G_L(w_1,\ld,w_K)\,{w_1\ov w_1\x-\t}\,dw_L\cd dw_1.\label{integral1}\ee
We do this by integrating out one variable at a time  until we get to the end. Precisely, we claim that for $k=1,\ld,L$ the integral equals

\be-{\x^{L-1}\ov (1-\x)^L}\,(\t^k/\x-1)^{L-k}\,{(1-\t)\cd(1-\t^{k-1})\ov \t^{kL }}\label{factor}\ee
\be\times \I0t\cd\I0t{(w_{k+1}-1)^{L-k-1}\ov w_{k+1}(w_{k+1}-\t)^{L-k}}\,\prod_{j\ge k+1}{\t/\x-w_j\ov \t^{k+1}/\x-w_j}\,G_{L-k}(w_{k+1},\ld,w_L)\,dw_L\cd dw_{k+1}.\label{integral2}\ee

For the inductive proof we use the easily checked fact that for $k\ge0$,
\be G_{L-k}(w_{k+1},\ld,w_L)={(w_{k+2}-1)^{L-k-2}\ov w_{k+2}(w_{k+2}-\t)^{L-k-1}}\,
\prod_{j>k+1}{w_{k+1}-w_j\ov\t w_{k+1}-w_j}\,G_{L-k-1}(w_{k+2},\ld,w_L).\label{G}\ee
Using Appendix B again we see that we can expand the $w_1$-contour in (\ref{integral1}) and find now that the integral equals minus the residue at $w_1=\t/\x$. Using (\ref{G}) with $k=0$ we find that (\ref{integral1}) equals the integral over $w_L,\ld,w_2$ of
\[-{(\t/\x-1)^{L-1}\ov \x\,(\t/\x-\t)^L}\,{(w_2-1)^{L-2}\ov w_2(w_2-\t)^{L-1}}\, \prod_{j\ge 2}{\t/\x-w_j\ov \t^2/\x-w_j}\,G_{L-1}(w_2,\ld, w_L)\]
\[=-{\x^{L-1}\ov\t^L\,(1-\x)^L}\,(\t/\x-1)^{L-1}\,{(w_2-1)^{L-2}\ov w_2(w_2-\t)^{L-1}}\,\prod_{j\ge 2}{\t/\x-w_j\ov \t^2/\x-w_j}\,G_{L-1}(w_2,\ld, w_L).\]

This verifies the claim for $k=1$. Now assume it is true for $k<L$. From Appendix~B, with $1$ replaced by $k+1$ and $L$ replaced by $L-k$, we find by expanding the contour that the integral over $w_{k+1}$ in (\ref{integral2}) equals minus the residue at $w_{k+1}=\t^{k+1}/\x$. Using (\ref{G}) we see that this equals the integral over $w_L,\ld,w_{k+2}$ of
\pagebreak
\[{(\t^{k+1}/\x-1)^{L-k-1}\ov \t^{k+1}/\x\;(\t^{k+1}/\x-\t)^{L-k}}\,(\t/\x-\t^{k+1}/\x)\,\,\prod_{j>k+1}{\t/\x-w_j\ov \t^{k+1}/\x-w_j}\]
\[\times\,{(w_{k+2}-1)^{L-k-2}\ov w_{k+2}(w_{k+2}-\t)^{L-k-1}}\,
\prod_{j>k+1}{\t^{k+1}/\x-w_j\ov \t^{k+2}/\x-w_j}\,G_{L-k-1}(w_{k+2},\ld,w_L)\]
\[={(\t^{k+1}/\x-1)^{L-k-1}\ov (\t^k/\x-\t)^{L-k}}\,{1-\t^k\ov\t^L}\]
\[\times\,{(w_{k+2}-1)^{L-k-2}\ov w_{k+2}(w_{k+2}-\t)^{L-k-1}}\,
\prod_{j>k+1}{\t/\x-w_j\ov \t^{k+2}/\x-w_j}\,G_{L-k-1}(w_{k+2},\ld,w_L).\]
If we multiply this by the factor (\ref{factor}) we obtain the statement for $k+1$. This completes the inductive proof of the claim.
When $k=L$ the integral in (\ref{integral2}) does not appear and we obtain the result that the integral (\ref{Fint}) equals
\be-{\x^{L-1}\ov(1-\x)^L}\,{(1-\t)\cd(1-\t^{L-1})\ov \t^{L^2}}.\label{result}\ee 

Putting all this together, we have shown that when we multiply $F(w_1,\ld,w_L)$ by (\ref{deteq}) and integrate the result is $c_3\inv\,{F_2}'(s)\, t^{-1/3}+o(t^{-1/3})$ times (\ref{result}). (Recall that (\ref{F0int}) equals zero, so the summand $F_2(s)+o(1)$ independent of the $w_j$ drops out, while (\ref{Fint}) gets multiplied by $c_3\inv\,{F_2}'(s)\,t^{-1/3}$ and $o(t^{-1/3})$ is added.)
If we combine this with (\ref{mint}) and refer to (\ref{PJw}) we obtain
\[\PP_{L,\,\Z^+}(x,m,t)=(-1)^{L-1}\,c_3\inv\,{\x^{L-1}\ov(1-\x)^L}\,{F_2}'(s)\,t^{-1/3}+o(t^{-1/3}).\]
Then using
\[\x=-\sqrt{\s}/(1-\sqrt{\s}),\ \ \ c_3=\s^{-1/6}\,(1-\sqrt{\s})^{5/3},\ \ \ c_2=\s^{-1/6}(1-\sqrt\s)^{2/3},\]
we see that
\[{c_3\inv\ov 1-\x}=c_2\inv,\ \ \ {\x\ov 1-\x}=-\sqrt\s.\]
This gives the statement of Theorem 1, that
\[\PP_{L,\,\Z^+}(x,m,t/\g)=c_2\inv\,\s^{(L-1)/2}\,{F_2}'(s)\,t^{-1/3}+o(t^{-1/3}).\]
\pagebreak

\bc{\bf V. Proof of Theorem 2}\ec

Let $P$ denote ASEP with step initial condition and $P'$ the process whose occupied sites are the unoccupied sites in $P$ after the switch $x\to -x$. Then $P'$ is an ASEP with step initial condition\footnote{Not quite since the initial configuration for $P'$ is the set of nonegative integers. This will make no difference and we ignore the fact.} and the same $p$ and $q$. 

We want to compute the probability $\PPb_{G,\,\Z^+}(x,m,t/\g)$ that $x_m=x$ in $P$ at time $t/\g$ and sites $x+1,\ld,x+G$ are unoccupied. In terms of $P'$ site $x'=-(x+G)$ is occupied, as are sites $x'+1,\ld,x'+G-1$, while site $x'+G=-x$ is unoccupied since $x$ is occupied in $P$. Thus, if $\PP'$ denotes probability for blocks in $P'$ then
\be\PPb_{G,\,\Z^+}(x,m,t/\g)=\PP_{G,\,\Z^+}'(x',m',t/\g)-\PP_{G+1,\,\Z^+}'(x',m',t/\g),\label{gaps}\ee
where $m'$ is such that $x_{m'}=x'$ in $P'$. To determine $m'$, observe that $m-1$ is the number of occupied sites in $P$ to the left of $x$, so analogously $m'-1$ is the number of unoccupied sites to the right of $x+G$ in $P$. The number of unoccupied sites to the right of $x$ equals $m-x$, so the number of unoccupied sites to the right of $x+G$ equals $m-x+O(1)$. Thus $m'=m-x+O(1)$.

The assumption of Theorem 1 is that we are in the KPZ regime in $P$, so that if $m=\s t$ then $\s$ is in a compact subset of $(0,1)$ and 
\be x=c_1\,t+c_2\,s\,t^{1/3}\ \ {\rm where}\ \ c_1=-1+2\sq,\ \ c_2=\sq^{\,-1/3}(1-\sq)^{2/3}.\label{x}\ee
To apply Theorem 1 to the right side of (\ref{gaps}) we have to show that we are in the KPZ regime in $P'$. This means that if we define
\be\s'=m'/t,\ \ c_1'=-1+2\sqp,\ \ c_2'=\sqp^{\,-1/3}(1-\sqp)^{2/3},\label{c'}\ee
then $\s'$ is in a compact subset of $(0,1)$ and
\be x'=c_1'\,t+c_2'\,s\,t^{1/3}+o(t^{1/3}).\footnote{It will turn out to be the same $s$. The $o(t^{1/3})$ doesn't matter since it can be removed by replacing~$s$ by $s+o(1)$, which will give the same asymptotics.}\label{x'}\ee
Since $m=\s t$ and $m'=m-x+O(1)$, it follows from (\ref{x}) that 
\[\s'=\s-c_1-c_2\,s\,t^{-2/3}+O(t\inv)=(1-\sq)^2-c_2\,s\,t^{-2/3}+O(t\inv),\]
\be\sqp=1-\sq-{1\ov2}(1-\sq)\inv c_2\,s\,t^{-2/3}+O(t\inv).\label{s'}\ee

Clearly $\s'$ is in a compact subset of $(0,1)$, so it remains to verify (\ref{x'}). 
First, from (\ref{c'}) and (\ref{s'}) it follows that
\[c_1'=-1+2\Big[1-\sq-{1\ov2}(1-\sq)\inv c_2\,s\,t^{-2/3}+O(t\inv)\Big]\]
\[=-c_1-(1-\sq)\inv c_2\,s\,t^{-2/3}+O(t\inv),\]
whence
\[ c_1't=-c_1t-(1-\sq)\inv c_2\,s\,t^{1/3}+O(1).\]
Then, from (\ref{c'}) and (\ref{s'}) again we see that
\[c_2'=(1-\sq)^{\,-1/3}\sq^{\,2/3}+O(t^{-2/3}),\]
so
\[ c_2'\,s\,t^{1/3}=(1-\sq)^{\,-1/3}\sq^{\,2/3}\,s\,t^{1/3}+O(t^{-1/3}).\]
Thus, 
\[c_1'\,t+c_2'\,s\,t^{1/3}=-c_1t-\Big[(1-\sq)\inv c_2-(1-\sq)^{\,-1/3}\sq^{\,2/3}\Big]\,s\,t^{1/3}+O(1).\]
Miraculously, the expression in brackets above equals $c_2$, as is easily checked. Therefore
\[c_1'\,t+c_2'\,s\,t^{1/3}=-c_1t-c_2\,s\,t^{1/3}+O(1).\]
Since $x'=-(x+G)=-c_1t-c_2st^{1/3}-G$, relation (\ref{x'}) follows. 

Hence we are in the KPZ regime in $P'$ and may apply Theorem 1 to the right side of (\ref{gaps}). We obtain
\[\PPb_{G,\,\Z^+}(x,m,t/\g)={c_2'}\inv\,(\s'^{\,(G-1)/2}-\s'^{\,G/2})\,{F_2}'(s)\,t^{-1/3}+o(t^{-1/3})\]
\[={c_2'}\inv\,(1-\sq)^{G-1}\,\sq\,{F_2}'(s)\,t^{-1/3}+o(t^{-1/3}).\]
Finally, one sees that ${c_2'}\inv\,\sq=c_2\inv\,(1-\sq)+O(t^{-2/3})$, so the above equals
\[c_2\inv\,(1-\sq)^{G}\,{F_2}'(s)\,t^{-1/3}+o(t^{-1/3}).\]
This completes the proof of Theorem 2.
\pagebreak

\bc{\bf Appendix A. Deformation to the operators \boldmath$J_{L,x,m}(w)$}\ec

In order to simplify notation, in this section we do not display the subscripts $L$ and $x$, since they will have the same meaning throughout. In particular the operator we denoted by $K_{L,x}(z)$ before we now denote by $K(z)$.

We follow the steps in the argument in \cite{tw3}. We first make the changes of variable
\[\x={1-\t\e\ov 1-\e},\ \ \x'={1-\t\e'\ov 1-\e'}, \ \ \ z_i={w_i-\t\ov w_i-1}\]
in the kernel of $K(z)$. The result is the operator $K_2(w)$ with kernel
\[K_2(\e,\e';w)={\phi(\e')\ov\e'-\t\e}\ \prod_{j=1}^L{w_j\,\e'-\t\ov w_j\,\e'-1},\]
where
\[\phi (\e)=\left(1-\t\e\ov1-\e\right)^{x+L-1}\,e^{[{1\ov1-\e}-{1\ov1-\t\e}]t}.\]
This operator, which has the same Fredholm determinant as $K(z)$, acts on functions on a small (clockwise) circle $\g$ about 1. Using
\[\prod_{j=1}^L {1\ov z_j^{L-j+1}(q z_j -p)}\prod_{i<j}{1\ov U(z_j,z_i)}\prod_j{dz_j\ov dw_j}=
(-1)^L q^{-L(L+1)/2}\prod_{j=1}^L{(w_j-1)^{L-j}\ov w_j (w_j-\t)^{L-j+1}}
\prod_{i<j}{w_j- w_i\ov w_j-\t w_i},\]
we see that (\ref{PLZ}) becomes
\[\PP_{L,\Z^+}(x,m,t)=-\t^{L(L+1)/2-(m+L-1)(L-1)}
\ \int_{\G_{0,\t}}\cd\int_{\G_{0,\t}} \prod_{j=1}^L{ (w_j-1)^{L-j}\ov w_j (w_j-\t)^{L-j+1}}\,
\prod_{i<j}{w_j-\t w_i\ov w_j-w_i}\]
\be\times \left[\int{\det(I-\la K_2(w))\ov(\t^L\la;\t)_m}\,{d\la\ov\la^L}\right]\, dw_L\cd dw_1,\label{K2int}\ee
where we made the substitution $\la\to \t^L\la$ in the $\la$-integral.  The $\la$-contour is large enough so that all the zeros of $(\t^L\la;\t)_{m}$ are inside the contour.

Now we shall use the two propositions on stability of Fredholm determinants stated in Section II. We introduce the operator $K_1(w)$ with kernel
\[ K_1(\e,\e';w)={\phi(\t\e)\ov\e'-\t\e}\,\prod_{j=1}^L {\t w_j\,\e-\t\ov\t w_j\,\e-1}.\]
Observe that $K_2(\e,\e';w)$ is analytic for $\e$ inside $\g$ while $K_1(\e,\e';w)$ is analytic for both $\e$ and $\e'$ inside $\g$. Therefore, by Proposition 2, the Fredholm determinant of $K_2(\e,\e';w)$ acting on $\g$ clockwise is equal to the Fredholm determinant of
\[ K_1(\e,\e';w)-K_2(\e,\e';w)\]
acting on $\g $ counterclockwise. Furthermore, if $\C$ denotes a (counterclockwise)  circle with center zero and radius $R\in(1,\t\inv)$, then the above difference is analytic in $\e$ and $\e'$ in the deformation from $\g$ to $\C$, the singularities at $\e'=\t\e$ of the two kernels cancelling. Using Proposition 1 we conclude that in the integral (\ref{K2int}) the operator $K_2(w)$ may be replaced by $K_1(w)-K_2(w)$ acting on $\C$.

Consider the kernel $K_1(\e,\e';w)$ acting on $r\C$, where $0\le r<1$. It is analytic for $\e,\,\e'\in r\C$ and so by Proposition 1 the Fredholm determinants are independent of $r$. On the other hand their Fredholm determinants are the same as for $r\,K_1(r\e,r\e';w)$ acting on $\C$. As $r\to0$ these operators converge in trace norm to $\t^L\,K_0$, where $K_0$ has kernel
\[K_0(\e,\e')={1\ov\e'-\t\e}.\footnote{The operator with kernel $r\,K_1(r\e,r\e';w)$ equals the trace class operator $K_0$ left-multiplied by a function tending uniformly on $\C$ to the constant function $\t^L$.}\]
We conclude that
\[\det(I-\la K_1(w))=\det(I-\t^L\la K_0)=(\t^L\la;\t)_\iy.\]
For the last identity see the end of the proof of Proposition 4 of \cite{tw3}.

Next, define $R(\la;w)=\la K_1(w)\,(I-\la K_1(w))\inv$, the resolvent operator of $K_1(w)$. From the preceding we conclude that $\det(I-\la K_2(w))$ appearing in (\ref{K2int}) is equal to
\[\det[I-\la K_1(w)+\la K_2(w)]=\det (I-\la K_1(w))\,\det[I+\la K_2(w) (I+R(\la;w)]\]
\[=(\t^L\la;\t)_{\iy}\, \det[I+\la K_2(w)\,(I+R(\la;w))],\]
the operators acting on functions on $\C$. Substituting this into (\ref{K2int}) we obtain
\[\PP_{L,\Z^+}(x,m,t)=-\t^{L(L+1)/2-(m+L-1)(L-1)}\int_{\G_{0,\t}}\cdots \int_{\G_{0,\t}} \prod_{j=1}^L{(w_j-1)^{L-j}\ov w_j (w_j-\t)^{L-j+1}}
\prod_{i<j}{w_j-\t w_i\ov w_j-w_i}\]
\be\times \left[\int (\la\t^{m+L};\t)_\iy \det\left[I+\la K_2(w) (I+R(\la;w))\right]
\,{d\la\ov\la^L}\right]\, dw_L\cdots dw_1.\label{P1}\ee

For the computation of $R(\e,\e';\la;w)$, the kernel of $R(\la;w)$, we can see by induction on $n$ that the kernel of $K_1(w)^n$ is 
\[\t^{nL}\,{\phi_n(\t\e)\ov \e'-\t^n\e}\,\prod_{j=1}^L{w_j\e-1\ov\t^n w_j\e-1},\]
where
\[\phi_n(\e)=\phi (\e)\phi (\t\e)\cd\phi(\t^{n-1}\e)=\left({1-\t^n\e\ov1-\e}\right)^{x+L-1}\,e^{[{1\ov1-\e}-{1\ov1-\t^n\e}]t}.\]
Therefore
\[R(\e,\e';\la;w)=\sum_{n=1}^\iy(\la\t^L)^n\,{\phi_n(\t\e)\ov \e'-\t^n\e}\,\prod_{j=1}^L{w_j\e-1\ov\t^n w_j\e-1}.\]
Since
\[\phi_n(\e)={\phi_\iy(\e)\ov\phi_\iy(\t^n\e)},\]
where
\[\phi_\iy(\e)=\lim_{n\to\iy}\phi_n(\e)=(1-\e)^{-x-L+1}\,e^{{\e\ov1-\e}t},\]
we can write the kernel of $K_2(w)R(\la;w)$ as
\[\sum_{n=1}^\iy (\t^L\la)^n\int {\phy(\z)\ov\phy(\t^{n+1}\z)}\,\prod_{j=1}^L {w_j\z-\t\ov\t^n w_j \z-1}\,{d\z\ov(\z-\t\e)(\e'-\t^n\z)}.\]
Here $|\z|=R$ but by analyticity we may take  any radius such that $1<|\z|<R$.
This is equal to
\[\sum_{n=1}^\iy (\t^L\la)^n\int {\phy(\z)\ov\z-\t\e} \prod_{j=1}^L (w_j\z-\t)\left[ \int{1\ov\phy(u\z)\,(\e'-u\z/\t)\,\prod_j (u w_j\z/\t-1)} \,{du\ov u-\t^{n+1}}  \right] \,d\z,\]
where on the $u$-contour we have $\t^2<|u|<\t R/|\z|$. (The $u$-integral equals the residue at $\t^{n+1}$.) Using
\[{1\ov u-\t^{n+1}}=\sum_{k=0}^\iy {\t^{(n+1)k}\ov u^{k+1}}\]
and summing first on $n$ gives
\[\sum_{k=0}^\iy {\la \t^{2k+L}\ov 1-\la \t^{k+L}} \int {\phy(\z)\ov \z-\t\eta} \prod_{j=1}^L (w_j\z-\t)\left[\int{1\ov\phy(u\z)\,(\eta'-u\z/\t)\,\prod_j (u w_j\z/\t-1)} \,{du\ov u^{k+1}}\right]\,d\z.\]

If we take $|u|>\t$ (as we may since $|\z|<R$) then when we write
\[{\la \t^{2k+L}\ov 1-\la \t^{k+L}}={\t^k\ov 1-\la \t^{k+L}}-\t^k,\]
the above sum may be written as the sum of two, in the obvious way, since both series will converge. Summing the second gives
\[-\int {\phy(\z)\ov \z-\t\eta} \prod_{j=1}^L (w_j\z-\t)\left[\int{1\ov\phy(u\z)\,(\eta'-u\z/\t)\,(u-\t)\,\prod_j (u w_j\z/\t-1)} \,du\right]\,d\z.\]
It is easy to check that the only singularity inside the $u$-contour is at $u=\t$, so evaluating the $u$-integral in the above gives
\[- \int {\phy(\z)\ov\phy(\t\z)}\,{1\ov(\z-\t\e)(\e'-\z)}\, \prod_{j=1}^L
{w_j\z-\t\ov w_j\z-1}\,d\z=- \int {\phi(\z)\ov(\z-\t\e)(\e'-\z)}\,
\prod_{j=1}^L{w_j\z-\t\ov w_j\z-1}\,d\z.\]
If we expand the contour so that $R<|\z|<\t\inv$, then we pass a pole at $\z=\eta'$ and get
\[ -{\phi(\e')\ov\e'-\t\e}\,\prod_{j=1}^L {w_j\eta'-\t\ov w_j\eta'-1} -  \int {\phi(\z)\ov(\z-\t\eta)\,(\eta'-\z)}\, \prod_{j=1}^L {w_j\z-\t\ov w_j\z-1}\,d\z.\]
The first term in the above is exactly $-K_2(\eta,\eta';w)$, so we have shown that 
\[(K_2(w)(I+R(\la;w))(\e,\e')=- \int{\phi(\z)\ov(\z-\t\e)(\e'-\z)} 
\prod_{j=1}^L {w_j\z-\t\ov w_j\z-1}\,d\z\]
\[+\sum_{k=0}^\iy {\t^{k}\ov1-\la \t^{k+L}} \int {\phy(\z)\ov\z-\t\e} \prod_{j=1}^L (w_j\z-\t)\left[\int {1\ov\phy(u\z)(\e'-u\z/\t)\prod_j (u w_j\z/\t-1)} \,{du\ov u^{k+1}}\right]\,d\z.\]
In the first integral on the right $|\z|>R$ while in the second $1<|\z|<R$ and 
$\t<|u|<\t R/|\z|$.

The first term on the right extends analytically for $\e,\,\e'$ inside $\C$ (the circle with radius~$R$) and the sum extends analytically for $\e$ inside $\C$. It follows by Proposition 2 that for the Fredholm determinant we may replace $K_2(w)(I+R(\la;w)$ by the operator whose kernel is the sum on the right. 

If in the sum the index $k$ were negative, then the $u$-integration would give zero since the integrand would be analytic inside the $u$-contour.  Therefore we may take the sum over $k\in\Z$. In addition we make the variable change $u\to u/\z$ and the sum becomes
\[\sum_{k\in\Z}{\t^{k}\ov1-\la \t^{k+L}} \int {\phy(\z)\ov\z-\t\e} \z^k\,\prod_{j=1}^L (w_j\z-\t)\left[\int {1\ov\phy(u)(\e'-u/\t)\prod_j (u w_j/\t-1)} \,{du\ov u^{k+1}}\right]\,d\z,\]
where now $\t|\z|<|u|<\t R$.

Recall that
\[f(\m,z)=\sum_{k\in\Z}{\t^k\ov1-\t^k\m}\,z^k,\]
which is analytic for $1<|z|<\t\inv$.
When we sum those terms involving $k$ we get
\[\sum_{k\in\Z}{\t^{k}\ov1-\la \t^{k+L}}\,(\z/u)^k=f(\la\t^L,\z/u),\]
so that the last sum equals
\[\t^L\int\int {\phy(\z)\ov\phy(u)}\,
 {f(\la\t^L,\z/u)\ov(\z-\t\eta)(\eta'-u/\t)}\,\prod_{j=1}^L {w_j\z-\t\ov u w_j-\t}\,
 d\z {du\ov u}.\]
 With the substitutions $\e,\e'\to \e/\t,\e'/\t$ we see that the above has the same Fredholm determinant as
\[\t^L\int\int {\phy(\z)\ov\phy(u)}\,
 {f(\la\t^L,\z/u)\ov(\z-\eta)(\eta'-u)}\,\prod_{j=1}^L {w_j\z-\t\ov u w_j-\t}\,
 d\z\,{du\ov u},\]
 where now the operator acts on a circle with radius $r\in(\t,1)$, and in the integrals $1<|\z|<r/\t,\ \t|\z|<|u|<r$.

We shall make the substitution $\la=\t^{-m}\m$ in (\ref{P1}). When we do that here we use the easy fact $f(\t^{-n}\m,z)=(\t z)^n\,f(\m,z)$ and see that the above becomes
\[\t^m\int\int {\phy(\z)\ov\phy(u)}\,\left({\z\ov u}\right)^{m-L}
{f(\m,\z/u)\ov(\z-\eta)(\eta'-u)}\,\prod_{j=1}^L {w_j\z-\t\ov u w_j-\t}\,
d\z \,{du\ov u}.\]
If we expand the $u$-contour, so that $r<|u|<1$ on the new contour, then we pass the pole at $u=\e'$ with minus the residue equal to
\[\t^m\int{\phy(\z)\ov\phy(\e')}\,{\z^{m-L}\ov (\e')^{m-L+1}}
{f(\m,\z/\e')\ov\z-\eta}\,\prod_{j=1}^L {w_j\z-\t\ov w_j\e'-\t}\,d\z=\t^m\,J_{L,x,m}(\e,\e';w),\]
by definition (\ref{Jkernel}). (The function $f(\m,\z/\e')$ remains analytic during the deformation.) The new double integral is anaytic for $|\e|,|\e'|\le r$ and $J_{L,x,m}(\e,\e';w)$ is analytic for $|\e|\le r$, so by Proposition 2 we may replace the last double integral by $J_{L,x,m}(\e,\e';w)$.
Making the substitution $\la=\t^{-m}\m$ in (\ref{P1}) and replacing $K_2(w) (I+R(\la;w))$ by $\t^m J _{L,x,m}(w)$ we arrive at (\ref{PJw}).
\pagebreak

\begin{center}{\bf Appendix B. Singularity of the \boldmath$G_L$-integral}\end{center}

Here we show that if $\ps(w_2,\ld,w_L)$ is analytic in the neighborhood of $\{0,\t\}^{L-1}$ then
\be\I0t\cd\I0t G_L(w_1,\ld,w_L)\,\ps(w_2,\ld,w_L)\,dw_L\cd dw_2\label{Gint}\ee
is analytic for $w_1$ outside $\{0,\t\}$ except for a pole of order at most $L-1$ at $w_1=1$, and is $O(1)$ for large $w_1$. This is straightforward except for the nature of the singularity $w_1=1$, which occurs when some $w_j$ is integrated over $\G_\t$. We show that it is pole of order at most $L-1$.

The integral (\ref{Gint}) is a sum of integrals in each of which some $w_j$ are taken over $\G_0$ and some over $\G_\t$. Let $Z$ be the set of former indices, and integrate first with respect to the $w_j$ with $j\in Z$. There are simple poles at $w_j=0$. When evaluating the integrals we recall the convention, which is relevant only for the integrals over $\G_0$, that we integrate first with respect to the $w_j$ with the largest $j$. If in the double product in (\ref{G0}) some $j\in Z$ then the product over $i$ equals $\t^{j-1}$, while if $i\in Z$ but $j\not\in Z$ then that factor becomes 1. Thus after these integrations we are left with a constant times (\ref{Gint}) except that all indices run over $Z^c$, and all integrals are over $\G_\t$. In the function $\ps(w_2,\ld,w_L)$  the variables $w_j$ with $j\in Z$ are set equal to zero.

To be more explicit, set $n=|Z|$ and let the indices in $Z^c$ be $\l_2<\l_3<\cd<\l_{L-n}$. (Note that $\l_2\ge 2$.)  Then the integrand becomes a function analytic near all $w_{\l_j}=\t$ and $w_1=1$ times
\[{\prod_{1<i<j}(w_{\l_i}-w_{\l_j})\ov 
\prod_{j>1}[(w_{\l_j}-\t)^{L-\l_j+1}(\t w_1-w_{\l_j})]}.\]
We make the variable changes $w_{\l_j}\to w_j+\t$, so the integrations are over $\G_0$ and the integrand equals is a function analytic near all $w_j=0$ ($j\ge2$) and $w_1=1$ times
\be{\prod_{1<i<j}(w_i-w_j)\ov \prod_{j>1}[w_j^{L-\l_j+1}(\t (w_1-1)-w_j)]}.\label{product}\ee 
It is conventient to set 
\[v=\t(w_1-1),\]
so that $\t (w_1-1)-w_j=v-w_j$. 
The analytic multiplying function has a series expansion
\be\sum_{\a_2,\ld,\a_{L-n}\ge0}c(\a_2,\ld,\a_{L-n})\prod_{j=2}^{L-n}w_j^{\a_j},\label{series}\ee
where the coefficients are polynomials in $v$. 

The numerator in (\ref{product}) is equal to a Vandermonde determinant $\det(w_j^{\; i})$. (Rows $i$ run from 0 to $L-n-2$ and columns $j$ from 2 to $L-n$.)
Multiplying by the rest of (\ref{product}) and the product in (\ref{series}) has the effect of multiplying column $j$ by $w_j^{-L+\l_j-1+\a_j}/(v-w_j)$. So we obtain a linear combination over $\a_2,\ld,\a_{L-n}$  of determinants with $i,j$ entry 
\be{w_j^{i-L+\l_j-1+\a_j}\ov v-w_j}.\label{entry}\ee
We are to integrate this over all $w_j\in\G_0$. Since $w_j$ appears only in column $j$, the multiple integral is gotten by integrating each entry. The $i,j$-entry integrates to $v^{i-L+\l_j-1+\a_j}$ if $i-L+\l_j+\a_j\le 0$, when the exponent is negative, and zero otherwise.\footnote{If any  $\a_j>L-\l_j$ then the $j$th column is zero, and therefore so is the determinant. Thus the linear combination is a finite one.}

We make the substitution $i\to L-n-2-i$ (the new $i$ also runs from 0 to $L-n-2$) and set $V=v\inv$. Then the $i,j$ entry is $V^{i+n+3-\l_j-\a_j}$ if the exponent is positive and zero otherwise. We factor out $V$ from each entry and get $V^{L-n-1}$ times the determinant whose $i,j$ entry is 
$V^{i+n+2-\l_j-\a_j}$
if the exponent is nonnegative and zero otherwise. Since for what follows we want $j$, as well as $i$, to start from 0 we replace the nonzero entries by
\be V^{i+n+2-\l_{j+2}-\a_{j+2}}.\label{entries}\ee

\noindent{\bf Lemma}. Suppose we have a determinant with $i,j$ entry $V^{a_i-b_j}$ if the exponent is nonnegative and zero otherwise. (The indices begin at 0.) If the determinant is nonzero then the $a_i$ and $b_j$, after reordering, satisfy
\[b_0\le a_0<b_1\le a_1<b_2\le a_2<\ldots,\]
in which case the determinant equals $\pm V^{\,\sum_i (a_i-b_i)}$.

\noindent{\bf Proof}. We may assume the sequences $\{a_i\}$ and $\{b_i\}$ are nondecreasing. If the determinant is nonzero then some exponent $a_0-b_j$ in the top row is nonnegative. Then this would hold for the smallest $b_j$, which is $b_0$. If also $a_0-b_j$ were nonnegative for some $j>0$ then columns $0$ and $j$ would be linearly dependent since there would be no zero entries in these columns. (Because the remaining $a_i\ge a_0$.) Hence $b_j>a_0$ for $j>0$ and the $0,0$-entry is the only nonzero one in the top row. Therefore the determinant equals $V^{a_0-b_0}$ times the $0,0$ cofactor, which is of the same form as the original determinant. The result follows by induction.

It follows from the lemma that when $a_i=i$ and the $b_j$ are integers the determinant is nonzero only if the $b_j$ when reordered satisfy $b_0\le0$ and $b_j=j$ when $j>0$, in which case the determinant is $\pm V^{-b_0}$. For the determinant with entries (\ref{entries}) we have $a_i=i$ and 
\[b_j=-n-2+\l_{j+2}+\a_{j+2}\ge-n\]
for all $j$ since $\a_{j+2}\ge0$ and $\l_{j+2}\ge2$. Thus the determinant is $O(V^n)$. Recalling the factor $V^{L-n-1}$ we had for this determinant we have shown that our original determinant is $O(V^{L-1})=O((w_1-1)^{-L+1})$. Therefore the singularity at $w_1=1$ is a pole of order at most $L-1$.

\bc{\bf Acknowledgments}\ec

The authors had interesting communications with Ivan Corwin who has found a way, from other considerations, to conjecture Corollaries 1 and 2. The authors also thank Timo Sepp\"al\"ainen for helpful comments.

The work was supported by the National Science Foundation through grants DMS--1207995 (first author) and DMS--1400248 (second author).

\end{document}